\documentclass[12pt]{article}

\usepackage{amsmath}
\usepackage{colonequals}
\usepackage{graphicx}
\usepackage{amssymb}
\usepackage{epstopdf}
\usepackage{graphicx}
\usepackage{subfigure}
\usepackage{pstricks-add}
\usepackage{pst-slpe}
\usepackage{color}
\usepackage{amsthm}
\usepackage{authblk}
 \usepackage[small]{caption}
\usepackage[bookmarks, colorlinks=true, linkcolor=blue, citecolor=green]{hyperref}

\title{Networks with the Smallest Average Distance and the Largest Average Clustering}
\author[1,2]{Dionysios Barmpoutis}
\author[2]{Richard M. Murray}
\affil[1]{Computation and Neural Systems}
\affil[2]{Control and Dynamical Systems \bigskip}
\affil[ ]{California Institute of Technology\bigskip}
\affil[ ]{dionysios$@$caltech.edu}
\affil[ ]{murray$@$cds.caltech.edu}
\date{\today}                                        

\newtheorem{lem}{Lemma}
\newtheorem{thm}{Theorem}
\newtheorem{cor}{Corollary}

\begin{document}
\maketitle
\begin{abstract}
We describe the structure of the graphs with the smallest average distance and the largest average clustering  given their order and size.
There is usually a unique graph with the largest average clustering, which at the same time has the smallest possible average distance. 
In contrast, there are many graphs with the same minimum average distance, ignoring their average clustering.
The form of these graphs is shown with analytical arguments.
Finally, we measure the sensitivity to rewiring of this architecture with respect to the clustering coefficient, and we devise a method to make these networks more robust with respect to vertex removal.
\end{abstract}
\section{Introduction}
Complex networks, as abstract models of large dynamical systems, match the structure of real-world networks in many diverse areas.
These include both natural and man-made systems such as gene regulation, protein interaction networks, food webs, economic and social networks and the internet, to name a few (see \cite{Strogatz2001} and references therein).
One of the prominent features that distinguishes them from random networks is the clustering among their individual units, quantified by their large clustering coefficient, and at the same time maintaining a small average path length among them.
These properties were thought to be mutually exclusive, but in almost all real networks, having a small average distance does not greatly interfere with the presence of a large clustering coefficient, as shown in \cite{WattsStrogatz1998}.
Despite the considerable amount of literature in the area of complex networks, there has been no study exploring the properties of a graph required in order for it to have both large average clustering and small average distance.
In this article, we will find the architecture of the networks with the largest average clustering, and show that at the same time their average distance is smaller or equal to the average distance of any other graph.
We will also study their resilience to vertex and edge removal, and how sensitive their properties are to edge rewiring.

\section{Preliminaries}
A graph is an ordered pair $\mathcal{G}=(\mathcal{V},\mathcal{E})$ comprising of a set $\mathcal{V}$ of vertices together with a set $\mathcal{E}$ of edges which are unordered 2-element subsets of $\mathcal{V}$.
Two vertices that are connected through an edge are called neighbors.
All graphs in this study are simple, meaning that all edges connect two different vertices, and there is no more than one edge between any two different vertices.
The order of a graph is the number of its vertices, $|\mathcal{V}|$.
A graph's size is $|\mathcal{E}|$ , the number of edges.
When there is no danger of confusion, we denote the graph with order $N$ and size $m$ as $\mathcal{G}(N,m)$.
A complete graph is a graph in which each vertex is connected to every other (with one edge between each pair of vertices).
The edge density of the graph is defined as $D=m/{N \choose 2}$, representing the number of present edges, as a fraction of the number of edges of a complete graph.
A clique in a graph is a subset of its vertices such that every two vertices in the subset are connected by an edge.
A clique that consists of three vertices (and three edges) is called a triangle.
The degree  of a vertex is the number of edges that connect to it.
A path is a sequence of consecutive edges in a graph and the length of the path is the number of edges traversed.
A path with no repeated vertices is called a simple path.
A graph is called connected if for every pair of vertices $u$ and $v$, there is a path from $u$ to $v$.
Otherwise the graph is called disconnected.
The distance $d(u,v)$ for a pair of vertices $u$ and $v$ in an undirected simple graph $\mathcal{G}$ is the shortest path from $u$ to $v$ (and vice versa).
A cycle is a closed (simple) path, with no other repeated vertices or edges other than the starting and ending vertices.
A cycle is called chordless when there is no edge joining two nodes that are not adjacent in the cycle.
A cut vertex of a connected graph is a vertex that if removed, (along with all edges incident with it) produces a graph that is disconnected.
A tree is a graph in which any two vertices are connected by exactly one simple path.
Finally, a subgraph $\mathcal{H}$ of a graph $\mathcal{G}$ is said to be induced if, for any pair of vertices $u$ and $v$ of $\mathcal{H}$, $(u,v)$ is an edge of $\mathcal{H}$ if and only if $(u,v)$ is an edge of $\mathcal{G}$.
 In other words, $\mathcal{H}$  is an induced subgraph of $\mathcal{G}$ if it has all the edges that appear in $\mathcal{G}$ over the same vertex set. 
If the vertex set of $\mathcal{H}$ is the subset $\mathcal{S}$ of $\mathcal{V}(\mathcal{G})$, then $\mathcal{H}$ can be written as $\mathcal{G}[\mathcal{S}]$ and is said to be induced by $\mathcal{S}$.

We will solve the problem of finding the graph which has the smallest average distance, and the largest average clustering coefficient, for given order $N$ and size $m$.

\section{Average Clustering and Average Distance}
The local clustering coefficient of a vertex $u$ is defined as the number of connections between vertices that are neighbors of $u$, divided by the total number of pairs of neighbors of $u$ \cite{WattsStrogatz1998}.
It is the number of triangles in which $u$ participates divided by the number of all possible triangles it could participate in, if all its neighbors were connected to each other.

More formally, if $d_{u}$ is the degree of a vertex $u$, and $t_{u}$ is the number of edges among its neighbors, its clustering coefficient is
\begin{equation}
C(u)= \left\{ \begin{array}{ll}
0 & \textrm{if $d_{u}=0$}\\
1 & \textrm{if $d_{u}=1$}\\
\frac{t_{u}}{{d_{u} \choose 2}} & \textrm{if $d_{u} \geq 2$}.
\end{array} \right.
\end{equation}
An example is shown in Figure \ref{DefinitionFigure1}.
The clustering coefficient of a vertex can only take values in the interval $[0,1]$.
Note that we deliberately choose to define the clustering coefficient of a vertex $u$ with degree $d_{u}=1$ as equal to $1$.
The graphs with the largest clustering under this convention may be different when we assume that vertices with degree $1$ have zero clustering.
The method to find the graphs with the largest clustering under the latter assumption is similar, and will be described later.

The average clustering coefficient for a graph $\mathcal{G}$ is simply the average of all the local clustering coefficients in its vertex set $\mathcal{V}(\mathcal{G})$.
A large average clustering coefficient is a proxy for increased robustness of the network, ``local'' structure and increased connection density among vertices in a neighborhood \cite{AshNewth2007}.
If $N$ is the order of the network, the average clustering coefficient is defined as
\begin{equation}
\bar{C}(\mathcal{G})=\frac{1}{N} \sum_{u \in \mathcal{V}(\mathcal{G})  } C(u).
\end{equation}
Since we will only be comparing graphs with the same number of vertices and edges, in order to make our analysis easier, we will be considering the sum of the clustering coefficients of all the vertices:
\begin{equation}
C_{S}(\mathcal{G})=\sum_{u \in \mathcal{V(G)} } C(u).
\end{equation}
Maximizing $C_{S}(\mathcal{G})$ is equivalent to maximizing $\bar{C}(\mathcal{G})$.
If a network has both a high average clustering coefficient, and a small average distance, it is called a \textit{ ``small world''} network \cite{WattsStrogatz1998}.
This architecture is conjectured to have other desired properties, like enhanced signal propagation speed, synchronizability and computational power \cite{WattsStrogatz1998},\cite{Strogatz2001}.
As it turns out, the networks with the largest average clustering are ``small world'' networks, since they also have the smallest possible average distance.

\begin{center}
\begin{figure}[htb]
\begin{center}
\subfigure[]{
\psscalebox{0.3}{
\begin{pspicture}(-5,-5)(7,7){
\cnodeput[fillstyle=solid,fillcolor=blue](1,4){A}{\strut}
\cnodeput(-3,2){B}{\strut}
\cnodeput(-3,-3){C}{\strut}
\cnodeput(1,-4){D}{\strut}
\cnodeput(4,0){E}{\strut}
\cnodeput(7,1.5){F}{\strut}
\cnodeput(6,5){G}{\strut}
}
\ncline{-}{A}{B}
\ncline{-}{A}{D}
\ncline{-}{A}{E}
\ncline{-}{A}{F}
\ncline{-}{A}{G}
\ncline{-}{B}{C}
\ncline[linecolor=red]{-}{B}{E}
\ncline[linecolor=red]{-}{B}{F}
\ncline{-}{C}{D}
\ncline{-}{C}{E}
\ncline[linecolor=red]{-}{D}{E}
\ncline[linecolor=red]{-}{F}{E}
\end{pspicture}
}
\label{DefinitionFigure1a}
}
\subfigure[]{
\psscalebox{0.3}{
\begin{pspicture}(-5,-5)(7,7){
\cnodeput(1,4){A}{\strut}
\cnodeput(-3,2){B}{\strut}
\cnodeput[fillstyle=solid,fillcolor=red](-3,-3){C}{\strut}
\cnodeput(1,-4){D}{\strut}
\cnodeput(4,0){E}{\strut}
\cnodeput(7,1.5){F}{\strut}
\cnodeput[fillstyle=solid,fillcolor=green](6,5){G}{\strut}
}
\ncline{-}{A}{B}
\ncline{-}{A}{D}
\ncline{-}{A}{E}
\ncline{-}{A}{F}
\ncline{-}{A}{G}
\ncline{-}{B}{C}
\ncline{-}{B}{E}
\ncline{-}{B}{F}
\ncline{-}{C}{D}
\ncline{-}{C}{G}
\ncline{-}{D}{E}
\ncline{-}{F}{E}
\end{pspicture}
}
\label{DefinitionFigure1b}
}
\subfigure[]{
\psscalebox{0.3}{
\begin{pspicture}(-2,0)(8,7){
\cnodeput[fillstyle=solid,fillcolor=blue](3.6,3.6){A}{\strut}
\cnodeput(0,3.6){B}{\strut}
\cnodeput(3.6,7.2){C}{\strut}
\cnodeput(0,7.2){D}{\strut}
\cnodeput(6.15,6.15){E}{\strut}
\cnodeput(8.69,3.6){F}{\strut}
\cnodeput(6.15,1.06){G}{\strut}
}
\ncline[linecolor=red]{-}{A}{B}
\ncline[linecolor=red]{-}{A}{C}
\ncline[linecolor=red]{-}{A}{D}
\ncline[linecolor=red]{-}{A}{E}
\ncline[linecolor=red]{-}{A}{F}
\ncline[linecolor=red]{-}{A}{G}
\ncline{-}{B}{C}
\ncline{-}{B}{D}
\ncline{-}{C}{D}
\ncline{-}{E}{F}
\ncline{-}{E}{G}
\ncline{-}{F}{G}
\end{pspicture}
}
\label{DefinitionFigure1c}
}
\end{center}
\caption{\textbf{(a)} The clustering coefficient of a vertex is the number of connections between its neighbors, divided by all pairs of neighbors, whether they are connected by an edge or not. The clustering coefficient of the blue vertex $u$ is $C(u)=4/{5 \choose 2}=0.4$. The average shortest path length is  $\bar{L}_{\mathcal{G}}=1.48$. \textbf{(b)} After rewiring one edge, and decreasing the distance between the red and the green vertex, the network has the smallest possible average distance, equal to 1.43. All vertex pairs now have at most a distance  of $2$.  \textbf{(c)} The network with the largest possible average clustering and smallest average shortest path length. The blue vertex is the central vertex of the induced star subgraph.}
\label{DefinitionFigure1}
\end{figure}
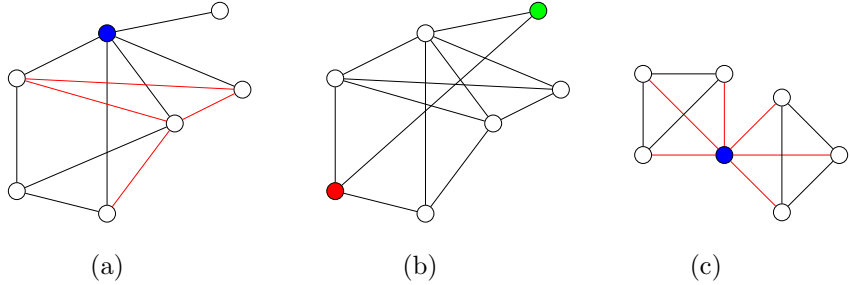
\end{center}
\section{Networks with the Smallest Average Distance}

The average distance of a network is an important property, since it is directly correlated with how different parts of the network communicate, and exchange information.
A small average distance is a proxy for improved synchronizability, efficient computation and signal propagation across the network \cite{Strogatz2001}.
In this section, we will analytically compute the minimum average distance of a graph of given order and size, and find a sufficient condition in order to achieve that minimum.

\begin{lem}
Assume we have a graph $\mathcal{G}(N,m)$ with average distance $\mathcal{L}_{\mathcal{G}}$, and we introduce one additional edge.
The graph $\mathcal{G'}(N,m+1)$ has an average distance $\mathcal{L}_{G'}<\mathcal{L}_{G}$.
\end{lem}
\begin{proof}
Assume that the edge connects two previously unconnected vertices $s$ and $t$, changing their distance to $d'(s,t)=1$.
Since they were not connected before, their distance was $d(s,t)\geq 2> d'(s,t)$.
For every other pair of vertices $u$ and $v$, $d'(u,v) \leq d(u,v)$, and hence, 
\begin{equation}
\mathcal{L}_{G'}=\frac{1}{{N \choose 2}}\sum _{\substack {(u,v) \in \mathcal{V}^{2}(\mathcal{G}) \\ u \neq v}} d'(u,v) < \frac{1}{{N \choose 2}}\sum _{\substack {(u,v) \in \mathcal{V}^{2}(\mathcal{G}) \\ u \neq v}} d(u,v) =\mathcal{L}_{G} .
\end{equation}
\end{proof}

\begin{lem}
The star graph is the only tree of order $N$ with the smallest average distance equal to $\mathcal{L}_{star}=2-\frac{2}{N} $.
\end{lem}
\begin{proof}
A tree has exactly $N-1$ edges among its $N$ vertices.
As a result, there will be exactly $N-1$ pairs of vertices with distance $d=1$, and  ${N-1 \choose 2}$ vertex pairs $(u,v)$ that are not connected, with distances $d(u,v) \geq 2$.
The star graph achieves this lower bound, and has the minimum possible average distance.
It is also unique. Assume that the tree is not a star, and as a result, there is no vertex that is connected to all the remaining vertices. In this case, there are at least two vertices with distance $d \geq 3$, since in every tree there is a unique path connecting each pair.
\end{proof}

Using the lemma above, we will find the smallest possible average distance of a graph with $N$ vertices and $m$ edges, which we denote $\mathcal{L}_{N,m}$.

\begin{thm}
The minimum possible average distance of a graph $\mathcal{G}(N,m)$  is equal to $\mathcal{L}_{N,m}=2-\frac{m}{{N \choose 2}} $.
\end{thm}
\begin{proof}
The graph $\mathcal{G}(N,m)$  has $m$ pairs of vertices with distance exactly $1$, and the rest of the pairs of vertices $(u,v)$ have distances $d(u,v) \geq 2$.
As a result, the minimum average distance is
\begin{equation}
\mathcal{L}_{\mathcal{G}} \geq \frac{m+ 2\left({N\choose 2} -m \right)}{{N \choose 2}}= 2-\frac{m}{{N \choose 2}}.
\label{lowestdistance}
\end{equation}
This lower bound can always be achieved.
A connected graph $\mathcal{G}(N,m)$ with a vertex $u$ of degree $d_{u}=N-1$ has an induced star graph, and as a result all non-neighboring vertices will have distance equal to $2$. All connected vertices have distance equal to $1$, leading to the lower bound of equation (\ref{lowestdistance}).
\end{proof}

\begin{cor}
If a graph $\mathcal{G}$ has two vertices $u$ and $v$ with distance $d(u,v) \geq 3$, then its average distance is $\mathcal{L}_{\mathcal{G}} >\mathcal{L}_{N,m}$.
\end{cor}
\begin{proof}
The number of pairs with distance $1$ is fixed, equal to the graph's size. 
All other vertices have a distance of at least $2$, and since the minimum is achieved when all non-neighboring pairs have a distance equal to the minimum,  the average distance of $\mathcal{G}$ is $\mathcal{L}_{\mathcal{G}} >\mathcal{L}_{N,m}$.
\end{proof}

It is worth noting that a graph $\mathcal{G}$ may have average distance $\mathcal{L}_{\mathcal{G}} =\mathcal{L}_{N,m}$ without having a star graph as an induced subgraph.
This is possible only if every pair of non-neighboring vertices has distance $d=2$.
An example is shown in Figures  \ref{DefinitionFigure1a} and \ref{DefinitionFigure1b}.
The blue and green vertices had distance equal to $3$ at first, and after rewiring one edge to connect them, the average distance became equal to the smallest possible.
It is easy to see that networks with the smallest distance need to have a star-like architecture.
Every pair of nodes needs to be part of at least one induced subgraph that is a star.

The above observations also propose an easy way to quickly decrease the average distance of a network by rewiring a small number of edges. 
At each step, we rewire an edge that is part of the most triangles, to connect the two vertices in the network with the largest distance.
The previously connected vertices have distance $2$, which is the smallest possible distance for unconnected vertices, and the newly connected vertices have the largest possible decrease in their distance.
Note that this method contributes to a large decrease in the average clustering of the network at the same time.
\section{Recursive Computation of Graph Clustering}

Assume that we add one vertex $u$ with degree $d$ to a graph $\mathcal{G}(N,m)$, by connecting it to the vertices in a set $\mathcal{D}$ ($|\mathcal{D}|=d$) and the result is a new graph $\mathcal{G}'(N+1,m+d)$.
The difference of the sum of clustering coefficients of the two graphs will be
\begin{equation}
\Delta C_{S}(\mathcal{G}',\mathcal{G})=C_{S}(\mathcal{G}')-C_{S}(\mathcal{G})=C(u)+\Delta C_{S}(\mathcal{G}',\mathcal{G},\mathcal{D})
\end{equation}
where $C(u)$ is the clustering coefficient of the new vertex, and 
\begin{equation}
\Delta C_{S}(\mathcal{G}',\mathcal{G},\mathcal{D})=\sum _{v \in \mathcal{D}} \left( C'(v)-C(v) \right)
\end{equation}
 is the sum of the differences of the clustering coefficients of the vertices in $\mathcal{D}$, before and after they acquire their new edge.
All other vertices have their clustering coefficient unchanged.
 
\begin{lem}
Assume that we have a graph $\mathcal{G}(N,m)$ with sum of clustering coefficients $C_{S}(\mathcal{G})$, and we add one more vertex $u$, with degree $d$.
The clustering difference of the two graphs  $\Delta C_{S}(\mathcal{G}',\mathcal{G})=C(u)+\Delta C_{S}(\mathcal{G}',\mathcal{G},\mathcal{D})$ can only be the largest possible if the vertices in $\mathcal{D}$ are part of a clique $Q$.
\end{lem}
\begin{proof}
If all vertices in $\mathcal{D}$ are part of a clique, they form an induced complete subgraph.
The clustering coefficient of $u$ will be the maximum possible ($C(u)=1$), since all the possible connections among its neighbors will be present.
Also, each of the $d$ vertices of $Q$ (with $|Q|=q\geq d$) that $u$ is connected to, will increase the number of connections among their neighbors to the maximum extent (given the degree of $u$), and they will have $d-1$ additional triangles each.
The clustering coefficient of the rest of the $q-d$ vertices of the clique will not be affected.
If, on the other hand, $u$ forms connections with vertices that do not form a clique, its clustering coefficient will be $C(u)<1$, and the vertices it is connected to will have less extra connections among their neighbors.
\end{proof}

Note that when $d<q$, the new clustering coefficient of the vertices that will be connected to $u$ will be less than $1$, possibly smaller than it was before connecting the new vertex.
In order to minimize the effect of the missing triangles on the overall clustering coefficient, we need to make sure that we connect $u$ to those vertices of the clique with the largest degree (which may have additional connections outside the clique).
Also, the size of the clique is an important factor.
The larger the clique, the smaller the impact of the additional vertex, since the degrees of the connected vertices are larger, but on the other hand we may have more missing triangles in total.

\begin{lem}
Assume that we have a graph $\mathcal{G}$ that consists of a clique $Q$ and two additional vertices $A$ and $B$ that each connects to a subset $\mathcal{D}_{1}$ and $\mathcal{D}_{2}$, with $d_{1}=|\mathcal{D}_{1}|\leq |\mathcal{D}_{2}|=d_{2} $.The average clustering of $\mathcal{G}$ is maximized when $\mathcal{D}_{1}\subseteq \mathcal{D}_{2}$.
\label{MergingLemma}
\end{lem}
\begin{proof}
The number of missing triangles stays the same regardless of the specific vertices of the clique that $A$ and $B$ are connected to.
So, the goal is to redistribute the missing triangles to vertices of the largest possible degree.
When $\mathcal{D}_{1}\subseteq \mathcal{D}_{2}$, the number of vertices of $Q$ with clustering equal to $1$ is maximized, and at the same time, the vertices of $Q$ with the largest degree have the maximum number of missing triangles.
\end{proof}
The above lemma can be used recursively for any number of edges connecting to a clique.
An example is shown in figure \ref{RedistributionTransformation}.
\begin{center}
\begin{figure}[htb]
\begin{center}
\subfigure[]{
\psscalebox{0.45}{
\begin{pspicture}(-6,-6)(7,8)
{
\cnodeput[fillstyle=solid,fillcolor=blue](1.24,3.8){A}{\strut}
\cnodeput[fillstyle=solid,fillcolor=orange](-3.24,2.35){B}{\strut}
\cnodeput(-3.24,-2.35){C}{\strut}
\cnodeput[fillstyle=solid,fillcolor=orange](1.24,-3.8){D}{\strut}
\cnodeput[fillstyle=solid,fillcolor=red](4,0){E}{\strut}
\cnodeput[fillstyle=solid,fillcolor=green](4,6){F}{\strut}
\cnodeput[fillstyle=solid,fillcolor=green](6,2){G}{\strut}
}
\ncline{-}{A}{B}
\ncline{-}{A}{C}
\ncline{-}{A}{D}
\ncline{-}{A}{E}
\ncline{-}{B}{C}
\ncline{-}{B}{D}
\ncline{-}{B}{E}
\ncline{-}{C}{D}
\ncline{-}{C}{E}
\ncline{-}{D}{E}
\ncline{-}{A}{F}
\ncline{-}{B}{F}
\ncline{-}{D}{F}
\ncline{-}{A}{G}
\ncline{-}{E}{G}
\end{pspicture}
}
\label{BeforeTransform1}
}
\subfigure[]{
\psscalebox{0.45}{
\begin{pspicture}(-6,-6)(7,8)
{
\cnodeput[fillstyle=solid,fillcolor=blue](1.24,3.8){A}{\strut}
\cnodeput(-3.24,2.35){B}{\strut}
\cnodeput(-3.24,-2.35){C}{\strut}
\cnodeput[fillstyle=solid,fillcolor=orange](1.24,-3.8){D}{\strut}
\cnodeput[fillstyle=solid,fillcolor=blue](4,0){E}{\strut}
\cnodeput[fillstyle=solid,fillcolor=green](3.5,5.3){F}{\strut}
\cnodeput[fillstyle=solid,fillcolor=green](6,1.6){G}{\strut}
}
\ncline{-}{A}{B}
\ncline{-}{A}{C}
\ncline{-}{A}{D}
\ncline{-}{A}{E}
\ncline{-}{B}{C}
\ncline{-}{B}{D}
\ncline{-}{B}{E}
\ncline{-}{C}{D}
\ncline{-}{C}{E}
\ncline{-}{D}{E}
\ncline{-}{A}{F}
\ncline{-}{D}{F}
\ncline{-}{E}{F}
\ncline{-}{A}{G}
\ncline{-}{E}{G}
\end{pspicture}
}
\label{AfterTransform1}
}
\end{center}
\caption{An example of a graph with two vertices connected to a clique. The graph in (a) has a smaller sum of clustering coefficients $C_{S}$ than the one in (b). By rewiring one edge in the graph on the left, we shift the missing triangles of one orange vertex to a blue one, with a higher degree. So the missing edges between a green and a white vertex have a smaller effect on the overall clustering.}
\label{RedistributionTransformation}
\end{figure}
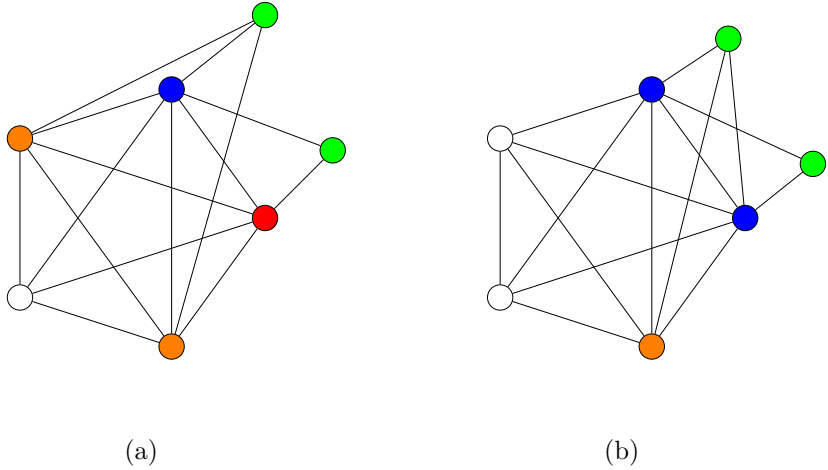
\end{center}

Assume we have a graph $\mathcal{G}$ which belongs to the set $\mathcal{S}_{N,m}$ of all graphs with $N$ vertices and $m$ edges.
We denote the clustering sum of the optimal graph with $C_{S}(N,m)$.
Its clustering sum will be
\begin{align}
C_{S}(N,m) &=\max _{\mathcal{G} \in \mathcal{S}_{N,m}}\left\{\sum_{u \in \mathcal{V}(\mathcal{G})} C(u) \right\} \\
&=\max _{\mathcal{G} \in \mathcal{S}_{N,m}} \left\{ \left( \sum_{u \in \mathcal{V}(\mathcal{G}) \backslash v} C(u)  \right) + C(v)\right\}
\label{SingleVertexRecursion}
\end{align}
for any vertex $v\in\mathcal{V}(\mathcal{G})$.
The reasoning behind the last equation is that if we pick any vertex from the graph, it should be ``optimally'' connected to a smaller graph that is itself optimal.
This means that we need to make sure that the chosen vertex has the largest possible clustering coefficient with regard to the vertices it is connected to, the rest of the graph needs to have the largest clustering coefficient, and the potential decrease of the clustering coefficient of the vertices after connecting the last one should be minimal.
This method provides an easy way to find the maximum clustering of a graph $\mathcal{G}$ of order $N$, by using a graph $\mathcal{G_{0}}$ of order $N-1$, and connecting one additional vertex to it.
The above can be algebraically expressed by conditioning on the set $\mathcal{D}$ of vertices with which the additional new vertex $v$ of degree $d=|\mathcal{D}|$ forms connections with.
\begin{equation}
C_{S}(N,m)=\max _{\mathcal{D} \subseteq \mathcal{V}(\mathcal{G}_{0})} \left\{C_{S}(N-1,m-d)+C(v) +\Delta C_{S}(N-1,m-d,\mathcal{D})\right\}
\label{RecursiveFormula}
\end{equation}
where $\Delta C_{S}(N-1,m-d,\mathcal{D})$ is the change in the clustering coefficient of the $d$ vertices of the set $\mathcal{D}$ of the graph of order $N-1$ and size $m-d$ when we connect one vertex (vertex $v$) with degree $d$ to them.
Again, the last equation argues that the graph with the largest possible clustering coefficient can be found by connecting a vertex to an optimal graph with fewer vertices and making sure that the algebraic value of the change is as large as possible.
\section{Almost Complete Graphs}
We call a connected graph of order $N$ and size $m$ \textit{almost complete} when its largest clique has order $N-1$  or $N-2$.
We distinguish these two cases by calling them \textit{type I} and \textit{type II}  respectively.
In order to be almost complete, a graph needs to have ${N-1\choose 2}+1 \leq m \leq {N\choose 2}-1$ (type I), and ${N-2\choose 2}+2 \leq m \leq{N-1\choose 2}$ (type II) edges.
The vertices of the largest clique are called \textit{central vertices}, whereas the vertices not belonging to it are called \textit{peripheral vertices}.
The two types of almost complete graphs are shown in Figure \ref{DefinitionFigure2}.\\
\begin{center}
\begin{figure}[htb]
	\begin{center}
	\subfigure[Type I]
	{
	\psscalebox{0.35}{
		\begin{pspicture}(-8,-6.5)(8,7)
		{
		\cnodeput(3.12,3.91){A}{\strut}
		\cnodeput(-1.11,4.87){B}{\strut}
		\cnodeput(-4.5,2.17){C}{\strut}
		\cnodeput(-4.5,-2.17){D}{\strut}
		\cnodeput(-1.11,-4.87){E}{\strut}
		\cnodeput(3.12,-3.91){F}{\strut}
		\cnodeput[fillstyle=solid,fillcolor=red](5,0){G}{\strut}
		}
		\ncline{-}{A}{B}
		\ncline{-}{A}{C}
		\ncline{-}{A}{D}
		\ncline{-}{A}{E}
		\ncline{-}{A}{F}
		\ncline{-}{B}{C}
		\ncline{-}{B}{D}
		\ncline{-}{B}{E}
		\ncline{-}{B}{F}
		\ncline{-}{C}{D}
		\ncline{-}{C}{E}
		\ncline{-}{C}{F}
		\ncline{-}{D}{E}
		\ncline{-}{D}{F}
		\ncline{-}{E}{F}
		\ncline{-}{A}{G}
		\ncline{-}{B}{G}
		\ncline{-}{C}{G}
	\end{pspicture}
		}	
	}	
	\subfigure[Type II]
	{
	\psscalebox{0.35}{		
		\begin{pspicture}(-8,-6)(8,7)
		{
		\cnodeput(1.24,3.8){A}{\strut}
		\cnodeput(-3.24,2.35){B}{\strut}
		\cnodeput(-3.24,-2.35){C}{\strut}
		\cnodeput(1.24,-3.8){D}{\strut}
		\cnodeput(4,0){E}{\strut}
		\cnodeput[fillstyle=solid,fillcolor=red](5.0347,6.5562){F}{\strut}
		\cnodeput[fillstyle=solid,fillcolor=red](7.7947,2.7562){G}{\strut}
		}
		\ncline{-}{A}{B}
		\ncline{-}{A}{C}
		\ncline{-}{A}{D}
		\ncline{-}{A}{E}
		\ncline{-}{B}{C}
		\ncline{-}{B}{D}
		\ncline{-}{B}{E}
		\ncline{-}{C}{D}
		\ncline{-}{C}{E}
		\ncline{-}{D}{E}
		\ncline{-}{F}{A}
		\ncline{-}{F}{E}
		\ncline{-}{G}{A}
		\ncline{-}{G}{E}
		\end{pspicture}
			}
		}
		\end{center}
\caption{\textbf{(a)} The type I almost complete graph consists of a clique of $N-1$ vertices, and one peripheral vertex (shown in red) that connects to them. \textbf{(b)} The type II almost complete graph of order $N$ consists of a clique of order $N-2$, and two additional vertices that connect to it (and possibly to each other).}
\label{DefinitionFigure2}
\end{figure}
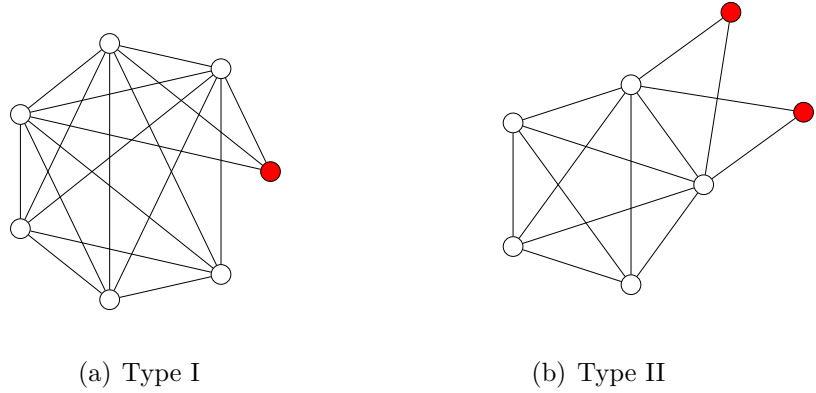
\end{center}

\begin{lem}
The total clustering for a type I almost complete graph $\mathcal{G}(N,m)$ is $C_{S}(\mathcal{G})=N-\frac{\alpha(N-1-\alpha)}{{N-1 \choose 2}}$, where $\alpha$ is the degree of the peripheral vertex.
\end{lem}
\begin{proof}
Assume that we have a type I almost complete graph as described above.
The peripheral vertex will have a clustering coefficient equal to $1$, the $\alpha$ vertices with degree $N-1$ will have a clustering coefficient $C(u)=\frac{{N-2 \choose 2}+(\alpha-1)}{{N-1 \choose 2}}$, and the $N-1-\alpha$ vertices with degree $N-2$ will have a clustering coefficient equal to 1 (see figure \ref{DefinitionFigure3a}).
The sum of the clustering coefficients for this graph will be
\begin{align}
C_{S}(\mathcal{G}) &=1+\alpha\frac{{N-2 \choose 2}+(\alpha-1)}{{N-1 \choose 2}}+(N-1-\alpha) \\
&=N+\frac{-\alpha {N-1 \choose 2}+\alpha{N-2 \choose 2}+\alpha(\alpha-1)}{{N-1 \choose 2}} \\
&=N+\frac{-\alpha(N-2)+\alpha(\alpha-1)}{{N-1 \choose 2}}  \\
&=N-\frac{\alpha(N-1-\alpha)}{{N-1 \choose 2}}.
\end{align}
\end{proof}
The average clustering, as a function of $\alpha$, is convex and symmetric around  $\frac{N-1}{2}$.
It decreases as $\alpha$ goes from $1$ to $\lfloor \frac{N-1}{2} \rfloor$, and then it increases as $\alpha$ goes from $\lceil \frac{N-1}{2} \rceil$ to $N-1$.

\begin{center}
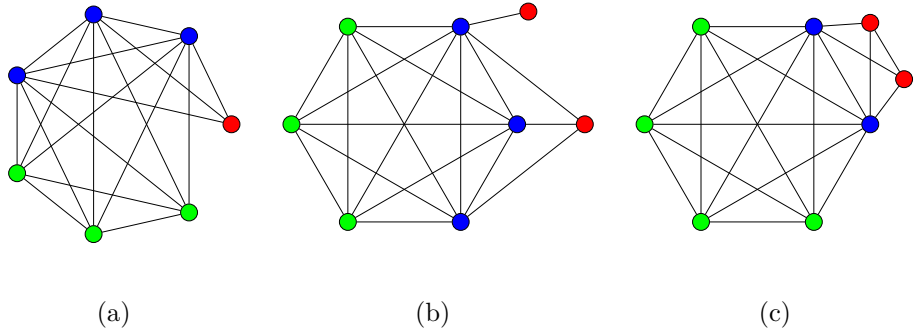
\begin{figure}[htb]
\centering
\subfigure[]{
\psscalebox{0.3}{		
\begin{pspicture}(-5.5,-7)(5.5,7)
{
\cnodeput[fillstyle=solid,fillcolor=blue](3.12,3.91){A}{\strut}
\cnodeput[fillstyle=solid,fillcolor=blue](-1.11,4.87){B}{\strut}
\cnodeput[fillstyle=solid,fillcolor=blue](-4.5,2.17){C}{\strut}
\cnodeput[fillstyle=solid,fillcolor=green](-4.5,-2.17){D}{\strut}
\cnodeput[fillstyle=solid,fillcolor=green](-1.11,-4.87){E}{\strut}
\cnodeput[fillstyle=solid,fillcolor=green](3.12,-3.91){F}{\strut}
\cnodeput[fillstyle=solid,fillcolor=red](5,0){G}{\strut}
}
\ncline{-}{A}{B}
\ncline{-}{A}{C}
\ncline{-}{A}{D}
\ncline{-}{A}{E}
\ncline{-}{A}{F}
\ncline{-}{B}{C}
\ncline{-}{B}{D}
\ncline{-}{B}{E}
\ncline{-}{B}{F}
\ncline{-}{C}{D}
\ncline{-}{C}{E}
\ncline{-}{C}{F}
\ncline{-}{D}{E}
\ncline{-}{D}{F}
\ncline{-}{E}{F}
\ncline{-}{A}{G}
\ncline{-}{B}{G}
\ncline{-}{C}{G}
\end{pspicture}
}
\label{DefinitionFigure3a}
}
\subfigure[]{
\psscalebox{0.3}{		
\begin{pspicture}(-5.5,-7)(8.5,7)
{
\cnodeput[fillstyle=solid,fillcolor=blue](2.5,4.33){A}{\strut}
\cnodeput[fillstyle=solid,fillcolor=green](-2.5,4.33){B}{\strut}
\cnodeput[fillstyle=solid,fillcolor=green](-5,0){C}{\strut}
\cnodeput[fillstyle=solid,fillcolor=green](-2.5,-4.33){D}{\strut}
\cnodeput[fillstyle=solid,fillcolor=blue](2.5,-4.33){E}{\strut}
\cnodeput[fillstyle=solid,fillcolor=blue](5,0){F}{\strut}
\cnodeput[fillstyle=solid,fillcolor=red](5.5,5){G}{\strut}
\cnodeput[fillstyle=solid,fillcolor=red](8,0){H}{\strut}
}
\ncline{-}{A}{B}
\ncline{-}{A}{C}
\ncline{-}{A}{D}
\ncline{-}{A}{E}
\ncline{-}{A}{F}
\ncline{-}{B}{C}
\ncline{-}{B}{D}
\ncline{-}{B}{E}
\ncline{-}{B}{F}
\ncline{-}{C}{D}
\ncline{-}{C}{E}
\ncline{-}{C}{F}
\ncline{-}{D}{E}
\ncline{-}{D}{F}
\ncline{-}{E}{F}
\ncline{-}{G}{A}
\ncline{-}{H}{E}
\ncline{-}{H}{A}
\ncline{-}{H}{F}
\end{pspicture}
}
\label{DefinitionFigure3b}
}
\subfigure[]{
\psscalebox{0.3}{		
\begin{pspicture}(-5.5,-7)(7.5,7)
{
\cnodeput[fillstyle=solid,fillcolor=blue](2.5,4.33){A}{\strut}
\cnodeput[fillstyle=solid,fillcolor=green](-2.5,4.33){B}{\strut}
\cnodeput[fillstyle=solid,fillcolor=green](-5,0){C}{\strut}
\cnodeput[fillstyle=solid,fillcolor=green](-2.5,-4.33){D}{\strut}
\cnodeput[fillstyle=solid,fillcolor=green](2.5,-4.33){E}{\strut}
\cnodeput[fillstyle=solid,fillcolor=blue](5,0){F}{\strut}
\cnodeput[fillstyle=solid,fillcolor=red](5,4.5){G}{\strut}
\cnodeput[fillstyle=solid,fillcolor=red](6.5,2){H}{\strut}
}
\ncline{-}{A}{B}
\ncline{-}{A}{C}
\ncline{-}{A}{D}
\ncline{-}{A}{E}
\ncline{-}{A}{F}
\ncline{-}{B}{C}
\ncline{-}{B}{D}
\ncline{-}{B}{E}
\ncline{-}{B}{F}
\ncline{-}{C}{D}
\ncline{-}{C}{E}
\ncline{-}{C}{F}
\ncline{-}{D}{E}
\ncline{-}{D}{F}
\ncline{-}{E}{F}
\ncline{-}{G}{A}
\ncline{-}{G}{F}
\ncline{-}{H}{A}
\ncline{-}{H}{F}
\ncline{-}{G}{H}
\end{pspicture}
}
\label{DefinitionFigure3c}
}
\caption{ Clustering for the almost complete graphs.  \textbf{(a)} The peripheral vertex (red) has unit clustering since it is connected to vertices of a clique. The vertices that are not connected to it (green) also have clustering equal to $1$, as members of the complete graph with no other edges. The vertices that are connected to the peripheral vertex (blue) have clustering less than $1$, since the peripheral vertex is not connected to all of their neighbors. \textbf{(b)} A type II almost complete graph with the largest clustering, where the two peripheral vertices are not connected (type IIa). \textbf{(c)} A maximum clustering coefficient type II almost complete graph, with connected peripheral vertices (type IIb). }
\label{TypeIAlmostCompleteGraphs}
\end{figure}
\end{center}

\begin{lem}
A type I almost complete graph  has larger clustering than any other nonisomorphic graph of the same order and size.
\end{lem}

\begin{proof}
Assume a type I almost complete graph $\mathcal{F}$ that consists of a clique $P$ of $N-1$ vertices, and one peripheral vertex of degree $\alpha$ that is connected to it.
$\mathcal{G}$ is a graph of the same order and size, whose largest clique $Q$ consists of $N-d \leq N-2$ vertices (otherwise it is isomorphic to $\mathcal{F}$).
The additional $d \geq 2$ vertices form connections to $Q$ and among each other.
Also in this case, we call these $d$ vertices  peripheral, and the $N-d$ vertices of $Q$ central.
Define $\gamma = {N \choose 2} -m$, which corresponds to the number of non-neighboring vertices in a graph.
By assumption, $1 \leq \gamma \leq N-2$.
It is easy to see that $d \leq \gamma$, since we have only $\gamma$ edges missing, and for each peripheral vertex $u$, at least $m_{u} \geq 1$ edges between itself and $Q$ have to be missing, otherwise $Q$ is not the largest clique.
We will add all of the clustering coefficients for all the vertices, and then show that the sum is always greater for $\mathcal{F}$.\\
We note the following:
\begin{itemize}
\item 
The number of vertices with clustering coefficient equal to $1$, is $1+\gamma$ for $\mathcal{F}$ and at most $d+\left\lfloor \frac{\gamma}{d} \right\rfloor$ for $\mathcal{G}$.
The number of such vertices is smaller or equal in $\mathcal{G}$ than in $\mathcal{F}$, for all $2 \leq d \leq \gamma$.
Conversely, $\mathcal{F}$ has exactly $N-1-\gamma$ vertices with clustering coefficient less than $1$, and $\mathcal{G}$, has at least $N-d-\left\lfloor \frac{\gamma}{d} \right\rfloor$ such vertices.
\item 
All vertices that have a clustering coefficient smaller than $1$ in $\mathcal{F}$ have degree $N-1$, the largest possible degree in a graph with $N$ vertices.

\end{itemize}
We will find the sum of the clustering coefficients for both graphs in terms of the number of triangles missing. 
Then, we will show that $C_{S}(\mathcal{F}) > C_{S}(\mathcal{G})$, for any $\mathcal{G} \neq \mathcal{F}$.

Since $\gamma=N-1-\alpha$, the sum of the clustering coefficients for $\mathcal{F}$, is
\begin{equation}
C_{S}(\mathcal{F})=N-\frac{\gamma (N-1-\gamma)}{{N-1\choose 2}}.
\end{equation}
The sum of the clustering coefficients for $\mathcal{G}$ is
\begin{align}
C_{S}(\mathcal{G}) & =N-\sum _{u=1}^{N} \frac{w_{u}}{{d_{u}\choose 2}} \\
                                  & \leq N-\frac{\sum _{u=1}^{N} w_{u}}{{N-1 \choose 2}}\\
                                  & \leq N-\frac{s_{g}}{{N-1 \choose 2}}
\end{align}
where $w_{u}$ is the number of edges missing among the neighbors of vertex $u$ and $s_{g}=\sum _{u=1}^{N} w_{u}$.

In order to prove the lemma, we only need to prove that  the missing triangles ($s_{g}$) of $\mathcal{G}$ are more than those  ($s_{f}$) missing in $\mathcal{F}$, in other words that
\begin{equation}
\Delta=s_{g}-s_{f}=\sum _{u=1}^{N} w_{u}-\gamma (N-1-\gamma)
\end{equation}
is greater than or equal to zero.

We will now find the minimum number of central vertices that are connected to at least two peripheral vertices.
This happens when $d-2$ peripheral vertices have the largest possible degree $N-d-1$, and the remaining two have the smallest possible degrees, and consequently the minimum number of clique vertices that they are both connected to.
If we assume that $b$ edges are \textit{not} present among peripheral vertices, we have a total of $t=d(N-d)-\gamma+b$ edges between peripheral vertices and central vertices.
If we further assume that all but $2$ peripheral vertices have the largest possible number of edges with the clique (which is $N-d-1$), the remaining peripheral vertices will have a degree sum of $r=t-(d-2)(N-d-1)=2N-d-2-\gamma+b$.
Since we only have $N-d$ vertices in $Q$, at least $r-(N-d)=N-2-\gamma+b$ of them will be connected to two peripheral vertices.

The total number of triangles missing from the central vertices, because of edges missing between peripheral and central vertices, is
\begin{align}
M_{C-P} &=\sum _{u=1}^{d} m_{u}(N-d-m_{u}) \\
&=(N-d) \left(  \sum _{u=1}^{d} m_{u} \right)  -\sum _{u=1}^{d} m_{u}^{2}.
\end{align}
The number of triangles missing because of the absence of edges among peripheral vertices, as shown above, is at least
\begin{equation}
M_{P-P} \geq \sum _{e=1}^{b} (N-2-\gamma+b)=b(N-2-\gamma+b).
\end{equation}

We are now ready to count the number of triangles missing from $\mathcal{G}$.
We omit the triangles missing from the peripheral vertices, and assume that all peripheral vertices have clustering equal to $1$ (otherwise the number of missing trinangles in $\mathcal{G}$ would further decrease).
The total number of triangles missing for $\mathcal{G}$ is
\begin{align}
s_{g} &\geq M_{C-P}+M_{P-P} \\
&\geq (N-d) \left(  \sum _{u=1}^{d} m_{u} \right)  -\sum _{u=1}^{d} m_{u}^{2} +b(N-2-\gamma+b) \\
&\geq (N-d)(\gamma -b)  -\sum _{u=1}^{d} m_{u}^{2} +b(N-2-\gamma+b)
\end{align}
with $\sum _{u=1}^{d} m_{u}=\gamma -b$, since $\gamma$ edges in total are missing.  
In order to minimize $s_{g}$, we need to maximize the sum of squares $\sum _{u=1}^{d} m_{u}^{2}$, under the constraints $m_{u} \geq 1$ for $u=1,2,\dots d$ and $\sum _{u=1}^{d} m_{u}=\gamma -b$.
The maximum is achieved when $m_{u}=1$ for $u=1,2,\dots d-1$, and $m_{d}=\gamma-b-(d-1)$.
As a result,
\begin{equation}
s_{g}\geq (N-d)(\gamma-b) -\left( (d-1)+(\gamma-b-d+1)^{2}\right) +b(N-2-\gamma+b)
\end{equation}
\begin{equation}
s_{g}\geq \gamma N-\gamma^{2}+\gamma d -bd+d+\gamma b-2\gamma-d^{2}.
\end{equation}
The difference between $s_{g}$ and $s_{f}$ is
\begin{align}
\Delta &=s_{g}-s_{f} \\
&\geq \gamma N-\gamma^{2}+\gamma d -bd+d+\gamma b-2\gamma-d^{2}-\gamma N +\gamma ^{2}+\gamma  \\
& \geq \gamma d -bd+d+\gamma b-\gamma-d^{2} \\
& \geq  (\gamma-d)(b+d-1).
\end{align}
By assumption, $d \geq 2$ and $b\geq 0$. 
The above product is always positive, when $d<\gamma$.
If $d=\gamma$, we have many  fully connected peripheral vertices, each of which connect to all but one vertex of $Q$.
No two peripheral vertices are connected to the same central ones, for in this case $Q$ would not be maximal.
As a result, there are less than $\gamma$ vertices in $Q$ that have unit clustering, and more than $N-1-\gamma$ vertices in $Q$ that are missing one or more triangles, and in that case too, $s_{g}>s_{f}$.
So in every case,  $C_{S}(\mathcal{F}) > C_{S}(\mathcal{G})$.
\end{proof}

There are two cases of type II almost complete graphs, depending on whether or not there is an edge between the two peripheral vertices.
If there is no such edge (type IIa), the graph with the largest clustering takes the form of Figure \ref{DefinitionFigure3b}.
When the two peripheral vertices are connected (type IIb), the graph with the form shown in Figure \ref{DefinitionFigure3c} has the largest average clustering coefficient, as shown in the next lemma.

\begin{lem}
Assume a type II almost complete graph $\mathcal{G}(N,m)$ with two peripheral vertices $u$ and $v$.
If $u$ and $v$ are not connected, the average clustering coefficient is maximized when $u$ and $v$ have the smallest number of common neighbors.
If $v$ and $v$ are not connected, the graph with the largest clustering coefficient is the one where they have the largest number of common neighbors.
\end{lem}
\begin{proof}
Let $c=m-{N-2 \choose 2}$ be the total number of edges connecting the two peripheral vertices to the rest of the graph.
Without loss of generality, assume that $u$ has smaller degree than $v$, $d(u)\leq d(v)$.
Then according to Lemma \ref{MergingLemma}, $c=d_{u}+d_{v}=2a+b$, where $a$ is the number of their common neighbors, and $b=d_{v}-d_{u}$ the neigbors of $v$ that are not connected to $u$.
The sum of clustering coefficients will be
\begin{align}
C_{s}(\mathcal{G}) &=N-a \cdot \frac{1+(N-2-a)+(N-2-a-b)}{{N-1 \choose 2}}-b\cdot \frac{N-2-a-b}{{N-2 \choose 2}}\\
&=N-\frac{c-b}{2} \cdot \frac{2N-3-c}{{N-1 \choose 2}}-b \cdot \frac{N-2-\frac{c-b}{2}-b}{{N-2 \choose 2}}.
\end{align}
Since $c$ is a constant, the sum of clustering coefficients is a function of $b$, and by differentiating, we find that it is increasing with $b$ for all $2 \leq c \leq N-2$ and $0 \leq b \leq c-2$.

Similarly, when  there is an edge between $u$ and $v$, $c=2a+b+1$, and the sum of vertex clustering coefficients is
\begin{align}
C_{s}(\mathcal{G}) &=N-a \cdot \frac{(N-2-a)+(N-2-a-b)}{{N-1 \choose 2}}-b\cdot \frac{N-2-a-b}{{N-2 \choose 2}}+\frac{b}{{a+b+1 \choose 2}}\\
&=N-\frac{c-b-1}{2} \cdot \frac{2N-3-c}{{N-1 \choose 2}}-b\cdot \frac{N-2-\frac{b+c-1}{2}}{{N-2 \choose 2}}+\frac{b}{{\frac{b+c+1}{2} \choose 2}}.
\label{Type2bAlmostCompleteAnalytical}
\end{align}
The last equation is a decreasing function of b for all $2\leq c \leq N-2$ and $0\leq b \leq c-3$.
From the above, we find that if a type II almost complete graph has maximum average clustering coefficient, there will be a peripheral vertex with degree $1$ (type IIa), or both vertices will have the maximum number of common neighbors (type IIb).
\end{proof}
For a given  size $m$ of a type II almost complete graph, we need to decide which of the two variations has the largest clustering.
We can easily see that the first two terms in both equations differ in the fact that $c$ and $c-1$ edges connect a peripheral with a central vertex respectively.
In the IIb type,  if $b \geq 1$, the third term in equation (\ref{Type2bAlmostCompleteAnalytical}) becomes very large, compared to the other terms for any $N>5$, and a necessary condition in order to have a type IIb almost complete graph is that $b=0$ (the two peripheral vertices have exactly the same neighbors), which means that $c$ has to be odd, $c=2a+1$.
Then, a simple comparison of the two equations shows that if 
\begin{equation}
c \geq  \frac{-1+3N+\sqrt{2 \left(12-N-4N^{2}+N^{3} \right)}}{1+N}
\end{equation}
then the type IIa has larger average clustering than type IIb.
The above number scales proportionally to the square root of the order $N$.

\begin{lem}
A type II almost complete graph has larger clustering than any other graph of the same order and size.
\end{lem}
\begin{proof}
We will use induction.
The claim is true for graphs of order $N=4$, shown by exhaustive enumeration of all the graphs with $4$ vertices.
Now, we will assume that it is true for all graphs of order up to $N-1$, and will show that this is still true for a graph of order $N$.
The optimal graph will be found by using equation (\ref{RecursiveFormula}).
The connected graph of order $N-1$ with the largest clustering coefficient is either a type I or type II almost complete graph for all possible degrees $d$ of the additional vertex $w$, because $1 \leq d \leq N-1$, and by assumption ${N-2\choose 2}+2 \leq m \leq {N-1 \choose 2}$, so
\begin{equation}
{N-3\choose 2} \leq m-d \leq  {N-1 \choose 2}-1.
\end{equation}
If it is a type I almost complete graph, addition of one vertex will transform it into an almost complete graph of type II, regardless of its degree.
Consequently, the lemma holds in this case.
Now assume that the existing graph is a type II almost complete graph.
Assume that the two already existing peripheral vertices $u$ and $v$ have $\alpha$ and $\beta$ edges with vertices of the largest clique.
If $\alpha+\beta+d=N-1$, then it is easy to see that the optimal graph consists of a full graph of order $N-2$, and two vertices of degree $1$ that connect to it. 
For every other value of $\alpha+\beta+d$,  since the initial graph is type II almost complete, $d\geq 2$, and $w$ has at least one neighbor in the clique.
Considering $u$, $v$ and $w$ in pairs, we can show that a graph with three peripheral vertices cannot be optimal, in other words, not all three can have less than $N-3$ edges to the clique of order $N-3$.
The reason is that according to equation \ref{SingleVertexRecursion}, any choice of a single vertex from the graph should yield the same result, in terms of maximizing the sum of clustering coefficients.
By the induction hypothesis, in each pair, one of the peripheral vertices should have  one connection to the clique, (not possible under the constraint $\alpha+\beta+d>N-1$) or all of them should be connected to each other and to the same vertices of the clique (also impossible for the same reason).
In all cases, a type II almost complete graph will have the largest average clustering coefficient.
\end{proof}

\section{Graphs with the Largest Clustering}

In this section, we will combine the previous results to show the form of the graphs with the largest clustering for a graph of arbitrary order and size.

\begin{lem}
The largest clustering graph with $N$ vertices and $0 \leq m \leq N-2$ edges consists of complete components of $2$ or $3$ vertices each.
\end{lem}
\begin{proof}
For m=0, we have no edges and the clustering coefficient is equal to zero.
If $m>0$, we first connect pairs of vertices, until all of them have degree $1$. 
If we have any edges left, we start forming triangles, by trying to keep the vertices that do not have any edges at a minimum.
In a triangle, we have the same number of vertices and edges, and since  $m \leq N-2$, the number of edges is not enough to connect all the vertices in triangles.
The above procedure will guarantee that the disconnected graph will have a sum of clustering coefficients equal to $1$.
\end{proof}

\begin{lem}
The tree that has the largest clustering (according to the convention that a vertex with degree $1$ has clustering $1$) is the star graph.
\end{lem}
\begin{proof}
Since there are no cycles in a tree (which also means no triangles), a vertex with degree larger than $1$ will have a clustering coefficient equal to zero.
By minimizing the number of such vertices (one vertex is the minimum number since the graph needs to be connected), we achieve the largest clustering for the star graph.
Also, note that the star graph is a graph with one cut vertex, connected to several complete graphs of order $2$.
\end{proof}

\begin{thm}
The graph with $N$ vertices and $N \leq m \leq {N-2 \choose 2}+1$ edges that has the largest possible clustering coefficient consists of one cut vertex that is shared by complete or almost complete subgraphs.
\end{thm}

\begin{proof}
We will use induction on the number of vertices to find the optimal graph for $N \leq m \leq {N-2 \choose 2}+1$.

For N=4, the statement is found to be true, by evaluating all the possible graphs.
(For computational considerations, see \cite{NetworkX,Software}. 
Assume that the optimal graph for every number of vertices up to $N-1$ and for the respective range of sizes, has the form mentioned above.

The graph with $N$ vertices and $m$ edges will be found by connecting a new vertex of degree $d$ to an optimal graph of order $N-1$ and size $m-d$.
The resulting graph will have the maximum possible clustering of the new vertex (equal to $1$), the maximum clustering of the rest of the graph it is attached to, and the minimal possible decrease in clustering for the vertices it is connected to.

If $d$ is larger than the order of the largest subgraph in $\mathcal{G}(N-1,m-d)$, then it will be connected to at least two smaller complete or almost complete subgraphs.
Now consider the subgraph that consists of these subgraphs plus the added vertex of degree $d$. 
It has $P<N$ vertices and $R<{P-2\choose 2}+2$ edges.
In addition, it has a cut vertex, which connects its two or more components.
This is a subgraph whose form is \textit{not} optimal, according to the induction hypothesis, meaning that there is a smaller degree $d$ for the vertex $u$, for which equation (\ref{RecursiveFormula}) gave a larger clustering coefficient for order $P$ and size $R$.
So we only need to consider vertices with degree $d$ less than the size of the largest ``module'' if $\mathcal{G}(N-1,m-d)$ is already connected.
Moreover, we only need to try to connect it to one clique.
The rest of the cliques will not change, so we focus on the clique where the new vertex with the new edges is added, and prove that it will still have the same form.
If it is a complete graph, then after adding the new vertex $u$, it will have one peripheral vertex, which makes the claim hold (the complete graph will now be a type II almost complete graph).
If it is a type I almost complete subgraph, after the addition of the new vertex, it will become a type II almost complete subgraph, as shown above.
If the subgraph is a type II almost complete subgraph, then connecting $u$ to it cannot yield an optimal subgraph, since by increasing the degree of $u$ to the size of this subgraph, we can make a graph with larger clustering, as shown in the lemma about the optimal form of type II almost complete graphs.
In any case, the new graph will have the form described in the claim.
\end{proof}

The arguments above just prove the form that the optimal graph needs to have, and not the exact arrangement of edges among its vertices.
Using equation (\ref{RecursiveFormula}), we can find the optimal graph in polynomial time, and as shown above, we only need to consider a small range of different degrees for the added vertex.

\begin{cor}
According to the above analysis, the graph $\mathcal{G}(N,m)$ with the largest possible clustering coefficient has one of the following forms, depending on its size $m$:
\end{cor}
\begin{itemize}
\item
$0 \leq m \leq N-2$: Disconnected graph, consisting of complete components with $2$ or $3$ vertices each.
\item
$N-1\leq m \leq {N-2 \choose 2}+1$: Complete or almost complete subgraphs that share one vertex.
\item 
${N-2 \choose 2}+2 \leq m \leq {N-1 \choose 2}$: Type II almost complete graph.
\item 
${N-1 \choose 2}+1 \leq m \leq {N \choose 2}-1$: Type I almost complete graph.
\item 
$m={N \choose 2}$: Complete graph, with  average clustering $\bar{C} \equiv 1$.
\end{itemize}

\section{Properties of the Graphs with the Largest Average Clustering Coefficient}
\begin{enumerate}
\item
The form of the optimal graphs is recursive.
Every combination of clusters (modules) that are connected through a cut vertex themselves form an optimal graph of the respective order and size.
This suggests an alternative way to generate an optimal graph, where it can be found as a combination of smaller complete or almost-complete subgraphs.
We can find the optimal form recursively with this procedure, too.
Since there is always a cut vertex, we can break the network in two smaller parts, that do not affect each other's average clustering coefficient, and optimize them independently.
\item
When a new vertex is attached to a graph with the process described above, it is \textit{always} attached to vertices that are members of the same clique.
This property can make computation easier, since we only need to consider cliques of size equal or greater to the degree of the candidate additional vertex, according to equation (\ref{RecursiveFormula}).
\item
No graph with the largest average clustering has any chordless cycles.
\item
No graph with the largest average clustering has any induced bipartite subgraphs.
\item
Every cut set is a complete graph of size $1$ or more.
In addition, if there exists a cut vertex, it also belongs to all larger cut sets.
\item
For all connected optimal graphs, there is at least one vertex that is connected to every other vertex in the graph.
If the graph is not almost complete, then the central vertex is unique.
Otherwise, we may have more than one such vertex.
\item
The networks with the largest clustering coefficient also have the smallest possible average distance among  the various vertices.
For every pair of vertices that are not connected, there is a path of length $2$ that connects them.
These networks are classified as ``small world'', because they have both high average clustering and small average shortest paths, as mentioned previously.
\item
From a computational standpoint, in order to find the optimal graph $\mathcal{G}(N,m)$, we first need to find the optimal graphs of order $N-1$, for all sizes, which in turn means that we need to compute \textit{all} such graphs of \textit{all orders up to} $N-1$.
The complexity of this procedure is polynomial,  $\mathcal{O}(N^{5})$, since we need to find the optimal graphs for all graphs of order up to $N$, each of which has $\mathcal{O}(N^{2})$ edges, and each time we need to add one vertex, trying various degrees $d=\mathcal{O}(N)$, to all appropriate $\mathcal{O}(N)$ cliques.
In order to avoid computing recursively the same optimal graphs, we need to start from the smallest possible optimal graph (for $N=3$), and build our way to the desired $N$, and storing all the optimal graphs to memory.
If we had to perform an exhaustive search, the complexity would be prohibitively large,  $\mathcal{O}(2^{N^{2}})$.
\end{enumerate}

\section{Fast Generation of Graphs with Small Distance and Large Average Clustering}

From the previous sections, it is now known that the form of the graphs with the smallest distance and largest clustering take the form of complete or almost complete subgraphs that are connected through one vertex.
In order to generate very large networks that have this form, we can resort to a much simpler algorithm that has almost constant complexity, and can generate networks with arbitrarily many vertices, with the minimal average distance and very close to optimal clustering coefficient.
Given the order and size of the graph, we find the largest complete subgraph, which leaves enough edges for the rest of the network to be connected.
Then we subtract the number of vertices and edges used, and repeat the process until all the vertices and edges have been used.
If at any point during the process, it is found that we cannot form a subgraph with the number of vertices and edges, we backtrack and reduce the cluster in the previous step (see \cite{Software}).
\section{Clustering Coefficient Sensitivity}
We have numerically computed the sensitivity of the network clustering under edge rewiring. 
In Figure \ref{ClusteringSensitivity}, we show the evolution of the average clustering coefficient after rewiring a percentage of the edges, for a typical large network.
If we rewire a small number of edges, this has no notable effect on the average clustering coefficient, since most of the vertices have only a small amount of neighbors that are not in their previous clique.
This is a similar situation as in \cite{WattsStrogatz1998}, with the only difference that the distance is very small from the start, and quite robust to rewirings, increasing less than $0.1\%$ even after we rewire a number of edges equal to the size of the network. 

\begin{figure}[htb]
   \centering
   \includegraphics[scale=0.4]{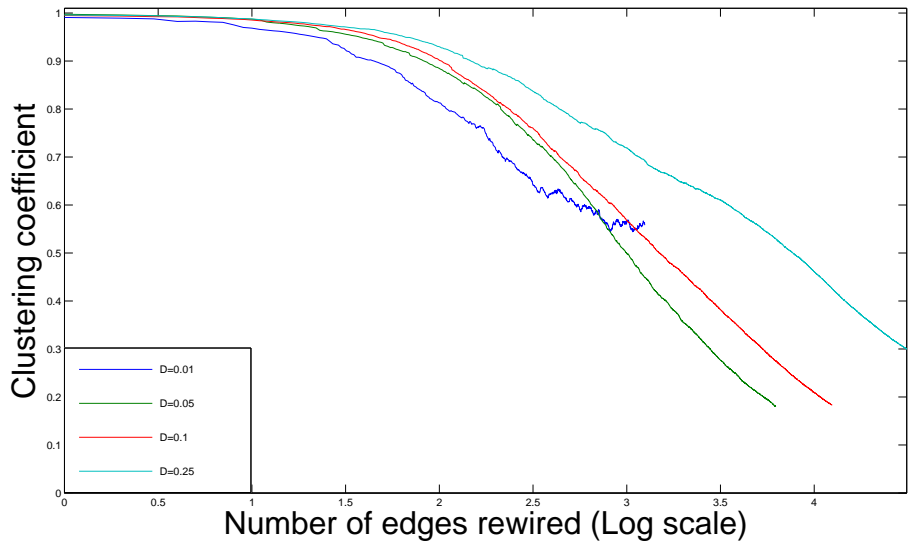} 
   \caption{Typical change of the clustering coefficient of a network with N=500 vertices and various edge densities $D$ . At every step one edge is rewired and the new average clustering coefficient is measured. We repeat the process for a number of steps equal to the number of edges in the graph. The rate of decrease in clustering is very small at first, but increases as we introduce more randomness in network architecture. Larger networks are more robust to rewiring a fixed number of edges.}
   \label{ClusteringSensitivity}
\end{figure}
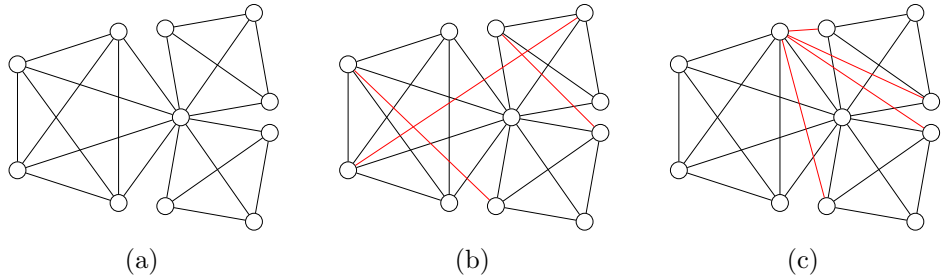
\begin{figure}[tbh]
\centering
\subfigure[]{
\psscalebox{0.3}{
\centering
\begin{pspicture}(-8,-5)(5,5)
{
\cnodeput(-2.76,3.8){A}{\strut}
\cnodeput(-7.24,2.35){B}{\strut}
\cnodeput(-7.24,-2.35){C}{\strut}
\cnodeput(-2.76,-3.8){D}{\strut}
\cnodeput(0,0){E}{\strut}
\cnodeput(-0.69,-3.94){F}{\strut}
\cnodeput(3.24,-4.63){G}{\strut}
\cnodeput(3.94,-0.69){H}{\strut}
\cnodeput(3.94,0.69){I}{\strut}
\cnodeput(3.24,4.63){J}{\strut}
\cnodeput(-0.69,3.94){K}{\strut}
}
\ncline{-}{A}{B}
\ncline{-}{A}{C}
\ncline{-}{A}{D}
\ncline{-}{A}{E}
\ncline{-}{B}{C}
\ncline{-}{B}{D}
\ncline{-}{B}{E}
\ncline{-}{C}{D}
\ncline{-}{C}{E}
\ncline{-}{D}{E}
\ncline{-}{E}{F}
\ncline{-}{E}{G}
\ncline{-}{E}{H}
\ncline{-}{F}{G}
\ncline{-}{F}{H}
\ncline{-}{G}{H}
\ncline{-}{E}{I}
\ncline{-}{E}{J}
\ncline{-}{E}{K}
\ncline{-}{I}{J}
\ncline{-}{I}{K}
\ncline{-}{J}{K}
\end{pspicture}
}
\label{NoAdditionalEdges}
}
\subfigure[]{
\psscalebox{0.3}{
\centering
\begin{pspicture}(-8,-5)(5,5)
{
\cnodeput(-2.76,3.8){A}{\strut}
\cnodeput(-7.24,2.35){B}{\strut}
\cnodeput(-7.24,-2.35){C}{\strut}
\cnodeput(-2.76,-3.8){D}{\strut}
\cnodeput(0,0){E}{\strut}
\cnodeput(-0.69,-3.94){F}{\strut}
\cnodeput(3.24,-4.63){G}{\strut}
\cnodeput(3.94,-0.69){H}{\strut}
\cnodeput(3.94,0.69){I}{\strut}
\cnodeput(3.24,4.63){J}{\strut}
\cnodeput(-0.69,3.94){K}{\strut}
}
\ncline{-}{A}{B}
\ncline{-}{A}{C}
\ncline{-}{A}{D}
\ncline{-}{A}{E}
\ncline{-}{B}{C}
\ncline{-}{B}{D}
\ncline{-}{B}{E}
\ncline{-}{C}{D}
\ncline{-}{C}{E}
\ncline{-}{D}{E}
\ncline{-}{E}{F}
\ncline{-}{E}{G}
\ncline{-}{E}{H}
\ncline{-}{F}{G}
\ncline{-}{F}{H}
\ncline{-}{G}{H}
\ncline{-}{E}{I}
\ncline{-}{E}{J}
\ncline{-}{E}{K}
\ncline{-}{I}{J}
\ncline{-}{I}{K}
\ncline{-}{J}{K}
\ncline[linecolor=red]{-}{B}{F}
\ncline[linecolor=red]{-}{C}{J}
\ncline[linecolor=red]{-}{H}{K}
\end{pspicture}
}
\label{DistributedAdditionalEdges}
}
\subfigure[]{
\psscalebox{0.3}{
\centering
\begin{pspicture}(-8,-5)(5,5)
{
\cnodeput(-2.76,3.8){A}{\strut}
\cnodeput(-7.24,2.35){B}{\strut}
\cnodeput(-7.24,-2.35){C}{\strut}
\cnodeput(-2.76,-3.8){D}{\strut}
\cnodeput(0,0){E}{\strut}
\cnodeput(-0.69,-3.94){F}{\strut}
\cnodeput(3.24,-4.63){G}{\strut}
\cnodeput(3.94,-0.69){H}{\strut}
\cnodeput(3.94,0.69){I}{\strut}
\cnodeput(3.24,4.63){J}{\strut}
\cnodeput(-0.69,3.94){K}{\strut}
}
\ncline{-}{A}{B}
\ncline{-}{A}{C}
\ncline{-}{A}{D}
\ncline{-}{A}{E}
\ncline{-}{B}{C}
\ncline{-}{B}{D}
\ncline{-}{B}{E}
\ncline{-}{C}{D}
\ncline{-}{C}{E}
\ncline{-}{D}{E}
\ncline{-}{E}{F}
\ncline{-}{E}{G}
\ncline{-}{E}{H}
\ncline{-}{F}{G}
\ncline{-}{F}{H}
\ncline{-}{G}{H}
\ncline{-}{E}{I}
\ncline{-}{E}{J}
\ncline{-}{E}{K}
\ncline{-}{I}{J}
\ncline{-}{I}{K}
\ncline{-}{J}{K}
\ncline[linecolor=red]{-}{A}{K}
\ncline[linecolor=red]{-}{A}{I}
\ncline[linecolor=red]{-}{A}{F}
\ncline[linecolor=red]{-}{A}{H}
\end{pspicture}
}
\label{CentralAdditionalEdges}
}
\caption{\textbf{(a)} The graph with the largest clustering coefficient of order $11$ and size $22$ consists of three complete subgraphs. Removal of the central vertex will render the network disconnected.\textbf{(b)} Addition of a few new edges among the different subgraphs creates new alternative communication paths among the various vertices of the network. As a result, the minimum cut set of the network is increased.\textbf{(c)} Adding new edges that connect one edge with vertices in the other subgraphs essentially creates a new network with more than one central vertex.}
\end{figure}

\section{Resilience to Vertex or Edge Removal}

The small-world networks studied here are very robust to edge removal.
Since almost every vertex is part of a complete graph, we need to remove at least as many edges as the order of the smallest clique it belongs to in order to render the network disconnected.
Even in that case, the number of vertices disconnected  is at most equal to the number of  edges deleted.

The situation is different for vertex removal.
We immediately note that unless we have enough edges to build an almost complete graph, the network becomes disconnected if we remove the central vertex, and the number of disconnected components is the number of modules in the initial network.
Real world networks on the other hand, rarely have an articulation point.
The robustness of a network to vertex failure is determined by its smallest cut set, and in this type of network, it consists of a single vertex (the central vertex), removal of which will render it disconnected.
Depending on the application, we may be able to add new edges, which will increase the network's robustness to vertex failure.
There are many ways to add the new edges, all of which result in reduced clustering.  
One of them is to distribute the new edges among vertices of the various modules, as shown in figure \ref{DistributedAdditionalEdges}.
This method ensures that if the central vertex is removed, there are still communication channels among the different subsystems of the network.
Another way is shown in figure  \ref{CentralAdditionalEdges}, where one or more of the vertices forms new connections with the vertices of all the other subgraphs, which in essence increases the number of 'central' vertices.
Rewiring  or adding new edges will have the least effect when they connect vertices with a large degree.
\begin{center}
\begin{figure}[tb]
	\subfigure[]
	{
	\psscalebox{0.31}{
	\begin{pspicture}(-1,-1)(12,8)
	{
	\cnodeput(0,4){A}{\strut}
	\cnodeput(0,0){B}{\strut}
	\cnodeput(4,0){C}{\strut}
	\cnodeput(4,4){D}{\strut}
	\cnodeput(8,0){E}{\strut}
	\cnodeput(12,0){F}{\strut}
	\cnodeput(12,4){G}{\strut}
	\cnodeput(8,4){H}{\strut}
	}
	\ncline{-}{A}{B}
	\ncline{-}{A}{C}
	\ncline{-}{A}{D}
	\ncline{-}{B}{C}
	\ncline{-}{B}{D}
	\ncline{-}{C}{D}
	\ncline{-}{C}{E}
	\ncline{-}{E}{F}
	\ncline{-}{E}{G}
	\ncline{-}{E}{H}
	\ncline{-}{F}{G}
	\ncline{-}{F}{H}
	\ncline{-}{G}{H}
	\end{pspicture}
		}
	\label{DefinitionFigure4a}
	}
	\subfigure[]
	{
	\psscalebox{0.28}{	
	\begin{pspicture}(-6,-6)(7,8)
	{
	\cnodeput(2.5,4.33){A}{\strut}
	\cnodeput(-2.5,4.33){B}{\strut}
	\cnodeput(-5,0){C}{\strut}
	\cnodeput(-2.5,-4.33){D}{\strut}
	\cnodeput(2.5,-4.33){E}{\strut}
	\cnodeput(5,0){F}{\strut}
	\cnodeput[fillstyle=solid,fillcolor=red](6.5,4){G}{\strut}
	\cnodeput[fillstyle=solid,fillcolor=red](7,1.5){H}{\strut}
	}
	\ncline{-}{A}{B}
	\ncline{-}{A}{C}
	\ncline{-}{A}{D}
	\ncline{-}{A}{E}
	\ncline{-}{A}{F}
	\ncline{-}{B}{C}
	\ncline{-}{B}{D}
	\ncline{-}{B}{E}
	\ncline{-}{B}{F}
	\ncline{-}{C}{D}
	\ncline{-}{C}{E}
	\ncline{-}{C}{F}
	\ncline{-}{D}{E}
	\ncline{-}{D}{F}
	\ncline{-}{E}{F}
	\ncline{-}{G}{A}
	\ncline{-}{G}{F}
	\ncline{-}{H}{A}
	\ncline{-}{H}{F}
	\end{pspicture}
			}
	\label{DefinitionFigure4b}
	}	
	\subfigure[]
	{
	\psscalebox{0.28}{		
	\begin{pspicture}(-7,-5)(5,8)
	{
	\cnodeput(3.54,3.54){A}{\strut}
	\cnodeput(0,5){B}{\strut}
	\cnodeput(-3.54,3.54){C}{\strut}
	\cnodeput(-5,0){D}{\strut}
	\cnodeput(-3.54,-3.54){E}{\strut}
	\cnodeput(0,-5){F}{\strut}
	\cnodeput(3.54,-3.54){G}{\strut}
	\cnodeput[fillstyle=solid,fillcolor=red](5,0){H}{\strut}
	}
	\ncline{-}{A}{B}
	\ncline{-}{A}{C}
	\ncline{-}{A}{D}
	\ncline{-}{A}{E}
	\ncline{-}{A}{F}
	\ncline{-}{B}{C}
	\ncline{-}{B}{D}
	\ncline{-}{B}{E}
	\ncline{-}{B}{F}
	\ncline{-}{C}{D}
	\ncline{-}{C}{E}
	\ncline{-}{C}{F}
	\ncline{-}{D}{E}
	\ncline{-}{D}{F}
	\ncline{-}{E}{F}
	\ncline{-}{G}{A}
	\ncline{-}{G}{B}
	\ncline{-}{G}{C}
	\ncline{-}{G}{D}
	\ncline{-}{G}{E}
	\ncline{-}{G}{F}
	\ncline{-}{H}{A}
	\ncline{-}{H}{G}
	\end{pspicture}
	}
	\label{DefinitionFigure4c}
	}	
\caption{Graphs with the largest clustering coefficient, defining the clustering coefficient of vertices with degree $d=1$ as equal to zero. \textbf{(a)} The optimal graph for $\mathcal{G}(8,13)$ has two cut vertices, connected through a bridge. \textbf{(b)} The smallest type II almost complete subgraph consists of a clique with $N-2$ vertices, and both peripheral vertices have degree at $2$. The peripheral vertices are shown in red. \textbf{(c)} The smallest type I almost complete graph under the new convention, where the peripheral vertex has degree $2$.}
\label{DefinitionFigure4}
\end{figure}
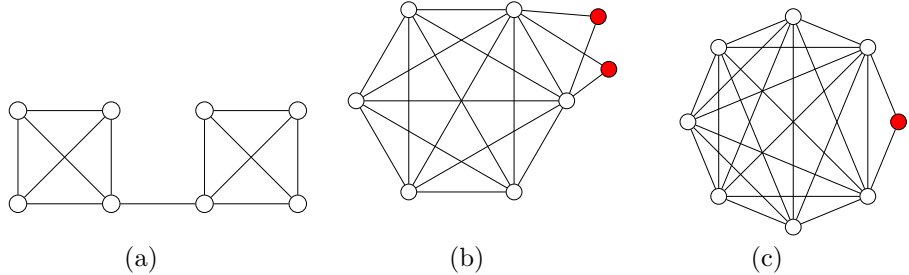
\end{center}

\section{Alternative Definition for Vertices with Degree $1$}

\begin{figure}[htbp]
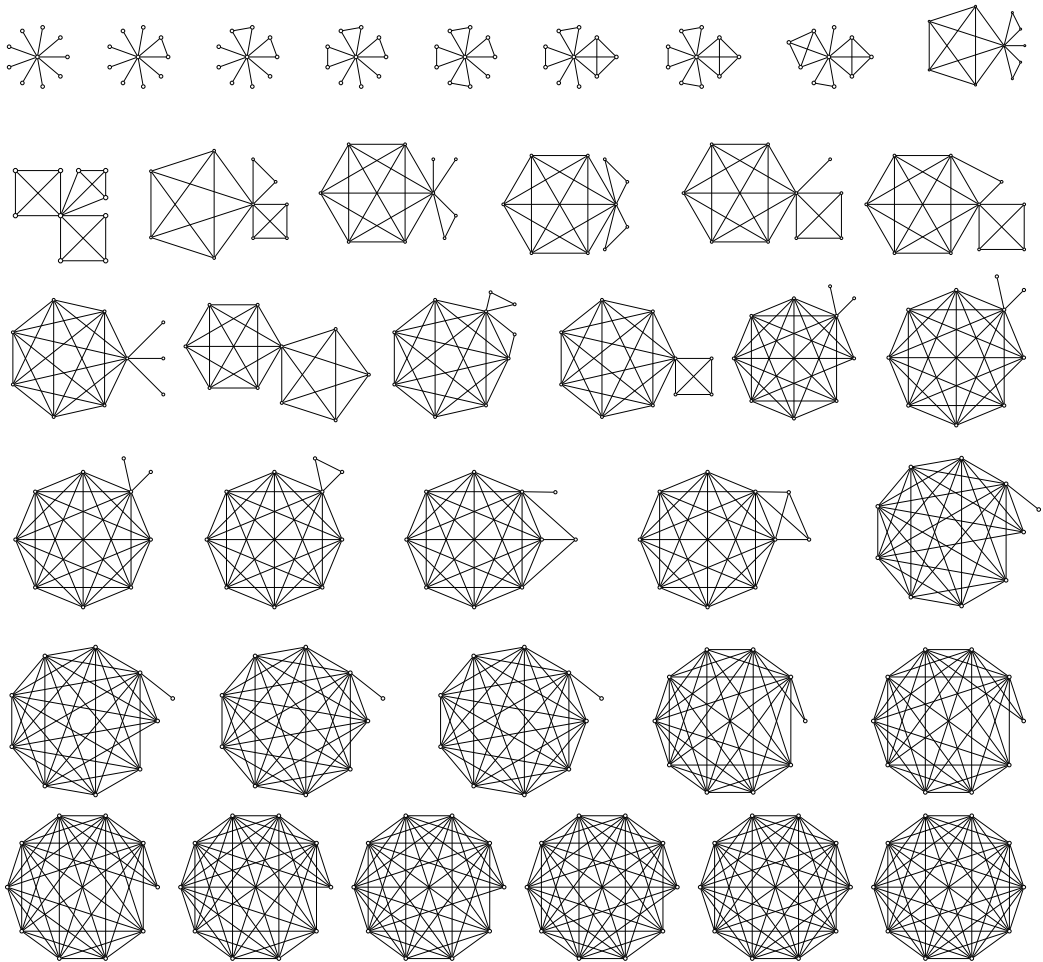

\subfigure{
\psscalebox{0.2}{

}
}
\caption{Graphs of order $10$ and size $9\leq m \leq 45$ with the largest clustering coefficient, assuming that vertices with degree $1$ have clustering coefficient equal to $1$.}
\label{All10New}
\end{figure}

\begin{figure}[htbp]
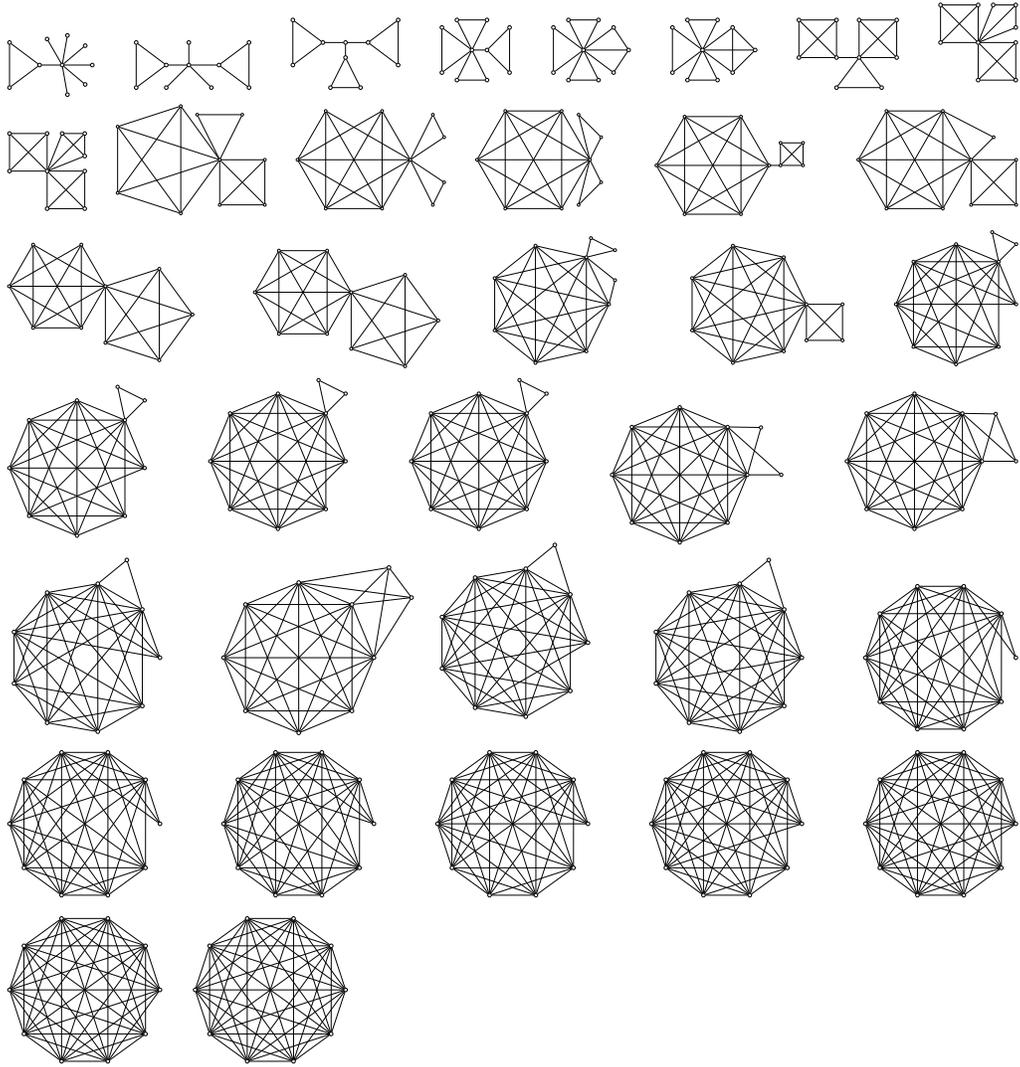

\subfigure{
\psscalebox{0.2}{
%
}
}
\caption{Graphs of order $10$ and size $10 \leq m \leq 45$ with the largest average clustering coefficient, assuming that vertices with degree $1$ have a clustering coefficient equal to zero.}
\label{All10Conventional}
\end{figure}

A common alternative convention for the vertices with degree $1$ is to define their clustering coefficient as equal to $0$.
The process of finding the graphs with the largest clustering under this new convention is very similar to the previous case.
The biggest difference is the form of graphs with a small number of edges, when they consist of at least one subgraph that is a tree.
For $m<N-1$, the graph is disconnected, and the optimal form consists of a group of disconnected triangles.
When $m=N-1$, the only form a connected graph can take is a tree, in which case, any arrangement of the vertices will yield a clustering coefficient of zero, since there can be no triangles.
As $m$ increases, we are able to form the first triangles, and the vertices that have nonzero clustering are part of a triangle, and have one more edge to the rest of the graph, keeping their degree low.
As the number of edges increases further, and every edge is part of at least one triangle, the form of the optimal graphs resembles the form under the previous case, with only one difference, and after which point, no vertex has degree $1$.
Some graphs, instead of having a unique cut vertex, have two cut vertices that are connected instead through one bridge edge, as shown in Figure \ref{DefinitionFigure4a}.
(Under the initial convention, the same graph would consist of the same two modules, plus one vertex with degree $1$, connected to the single cut vertex).
The last difference is that the definition of the almost complete graphs has to be changed so that the peripheral vertices have a degree of at least $2$ (Figures \ref{DefinitionFigure4b} and \ref{DefinitionFigure4c}.
The process of finding the optimal graphs remains otherwise the same (see \cite{Software}), and an example for a graph of order $10$ is shown in Figure \ref{All10Conventional}.
Comparing figures \ref{All10New} and  \ref{All10Conventional}, we immediately see the similarity of the optimal graphs in both cases.


\clearpage
\newpage
\section*{Acknowledgements}
This work was supported by the Institute for Collaborative Biotechnologies through contract number W911NF-09-D-0001 from the U.S. Army Research Office and by the Network Science and Engineering Project under NSF award CNS-0911041.


\begin{thebibliography}{10}
\bibitem{WattsStrogatz1998} Watts, DJ and Strogatz, SH, 
{ \it Collective Dynamics of 'small world' networks}, Nature {\bf 393},440--442, (1998).

\bibitem{AshNewth2007} Ash, J. and Newth, D. {\it Optimizing complex networks for resilience against cascading failure}, Physica A: Statistical Mechanics and its Applications {\bf 380}, 673-683 (2007).

\bibitem{Strogatz2001} Strogatz, SH {\it Exploring Complex Networks}, Nature {\bf 410},268--276, (2001).

\bibitem{NetworkX} Hagberg, AA Schult, DA and Swart, PJ {\it Exploring network structure, dynamics, and function using NetworkX},  Proceedings of the 7th Python in Science Conference, Gael Varoquaux, Travis Vaught, and Jarrod Millman (Eds), pp. 11-15, (2008)

\bibitem{Software} A Python program that implements this function is available from the authors upon request.

\end{thebibliography}
\end{document}